\documentclass[printer]{aa} % for a referee version
\usepackage{graphicx}
\usepackage{txfonts,natbib}

\def\gx{\mbox{GX 339-4}}

\def\xtejodh{\mbox{XTE J1118+480}}
\def\xtejqcq{\mbox{XTE J1550-564}}

\def\cmmoinsdeux{\mbox{ cm}^{-2}}

\def\microns{\mbox{ } \mu \mbox{m}}

\def\kpc{\mbox{ kpc}}

\def\Msol{\mbox{ }M_{\odot}}

\def\mags{\mbox{ magnitudes}}

\def\keV{\mbox{ keV}}

\def\amin{^\prime}
\def\aminp{{\rlap.}^{\prime}}
\def\asec{^{\prime \prime}}
\def\asecp{{\rlap.}^{\prime \prime}}

\def\Av{A_{\rm v}}

\def\nh{N_{\rm H}}

\def\ltsima{\; \buildrel < \over \sim \;}
\def\simlt{\lower.5ex\hbox{\ltsima}}            % < over MMM
\def\gtsima{\; \buildrel > \over \sim \;}
\def\simgt{\lower.5ex\hbox{\gtsima}}            % > over MMM

\begin{document}
\title{Near-infrared jet emission in the microquasar $\xtejqcq$\thanks{Based on observations collected at the European Southern Observatory, Chile, through programs 071.D-0071 and 079.D-0623.}
}

%   \subtitle{I. xxx}

\author{S. Chaty\inst{1} \and G. Dubus\inst{2} \and A. Raichoor\inst{1,3}}

   \institute{Laboratoire AIM (UMR 7158 CEA/DSM-CNRS-Universit\'e Paris Diderot), Irfu/Service d'Astrophysique, CEA-Saclay, FR-91191 Gif-sur-Yvette Cedex, France, \email{chaty@cea.fr}
   \and 
   Laboratoire d'Astrophysique de Grenoble, UMR 5571 Universit\'e Joseph Fourier Grenoble I / CNRS, BP 53, 38041 Grenoble, France %\email{Guillaume.Dubus@obs.ujf-grenoble.fr}
   \and GEPI, Observatoire de Paris-Meudon, 5 place Jules Janssen, 92195 Meudon, France %\email{anand.raichoor@obspm.fr}
}

   \date{Received August 15, 2010; accepted February 15, 2011}

% \abstract{}{}{}{}{} 
% 5 {} token are mandatory
 
  \abstract
   {Microquasars are accreting Galactic sources that are also observed to launch relativistic jets. A key signature of the ejection is non-thermal radio emission. The level of this jet component at high frequencies is still poorly constrained.}
   {The X-ray binary and microquasar black hole candidate $\xtejqcq$ exhibited a faint X-ray outburst in April 2003 during which it stayed in the  X-ray low/hard state. We took optical and near-infrared (NIR) observations with the ESO/NTT telescope during this outburst to distinguish the various contributions to the spectral energy distribution (SED) and investigate the presence of a jet component.} 
{Photometric and spectroscopic observations allowed us to construct an SED and also to produce a high time-resolution lightcurve.}
   {The SED shows an abrupt change of slope from the NIR domain to the optical. The NIR emission is attributed to non-thermal synchrotron emission from the compact, self-absorbed jet that is known to be present in the low/hard state. 
This is corroborated by the fast variability, colours, lack of prominent spectral features and evidence for intrinsic polarisation. The SED suggests the jet break from the optically thick to the thin regime occurs in the NIR.}
  {The simultaneous optical-NIR data allow an independent confirmation of jet emission in the NIR. The transition to optically thin synchrotron occurs at NIR frequencies or below, which leads to an estimated characteristic size $\ga 2\times 10^8$ cm and magnetic field $\la 5$\,T for the jet base, assuming a homogeneous one-zone synchrotron model.}


\authorrunning{S. Chaty et al.}
\titlerunning{NIR jet emission in the microquasar $\xtejqcq$}

   \maketitle
%
%________________________________________________________________

\section{Introduction} \label{introduction}

   X-ray binary systems are composed of a companion star and a compact
object \--- a black hole or a neutron star. 
In low-mass X-ray binaries (LMXBs), the companion star is a late-type star filling its
Roche lobe.
Matter transiting through the Lagrange point
forms an accretion disk around the compact object. The LMXBs spend most of their time in a quiescent state
with a low X-ray luminosity. Outbursts occasionally occur, owing to an instability in the accretion disk,  during which the
X-ray luminosity increases by several orders of magnitude. Those LMXBs that additionally show non-thermal radio emission that is sometimes spatially resolved into jets
 are called microquasars (see e.g. \citeauthor{chaty:2006a} \citeyear{chaty:2006a}; \citeauthor{chaty:2006b} \citeyear{chaty:2006b}; \citeauthor{fender:2006} \citeyear{fender:2006}; \citeauthor{mirabel:1998b} \citeyear{mirabel:1998b}).

Several canonical states for LMXBs have been defined according to their X-ray emission properties (see e.g. \citeauthor{belloni:2010} \citeyear{belloni:2010}; \citeauthor{remillard:2006} \citeyear{remillard:2006}), the main ones being:

\begin{itemize}

\item {\it the high/soft state}, characterized in the X-rays by a high luminosity, dominated by the thermal emission of the accretion disk, with a peak temperature $\sim 1-1.5 \keV$, emitting as a multicolour blackbody from optical to X-rays,

\item {\it the low/hard state}, with the X-rays dominated by a power-law component, 
the accretion disk being weak in the X-ray band, at a temperature $\sim 0.01-0.5 \keV$; 
the hard state is invariably associated with strong, flat spectrum radio emission that is attributed to a compact, self-absorbed jet (with an extension of $\sim 10^{-6}$\,$\asec$, see e.g. \citeauthor{fender:2006} \citeyear{fender:2006}).

\item {\it the quiescent state}, when the X-ray luminosity is low and the optical/infrared emission is dominated by the emission of the companion star.

\end{itemize}

We also point out the existence of {\it the intermediate state}. During the transition between the low/hard and the high/soft state, discrete ejections are most of the time observed in radio, with an extension $\sim 0.1-10 \asec$ (see e.g. \citeauthor{fender:2006}  \citeyear{fender:2006}). \\

\begin{figure*}[!ht]
	\begin{center}
	\includegraphics[angle=-90,width=9cm]{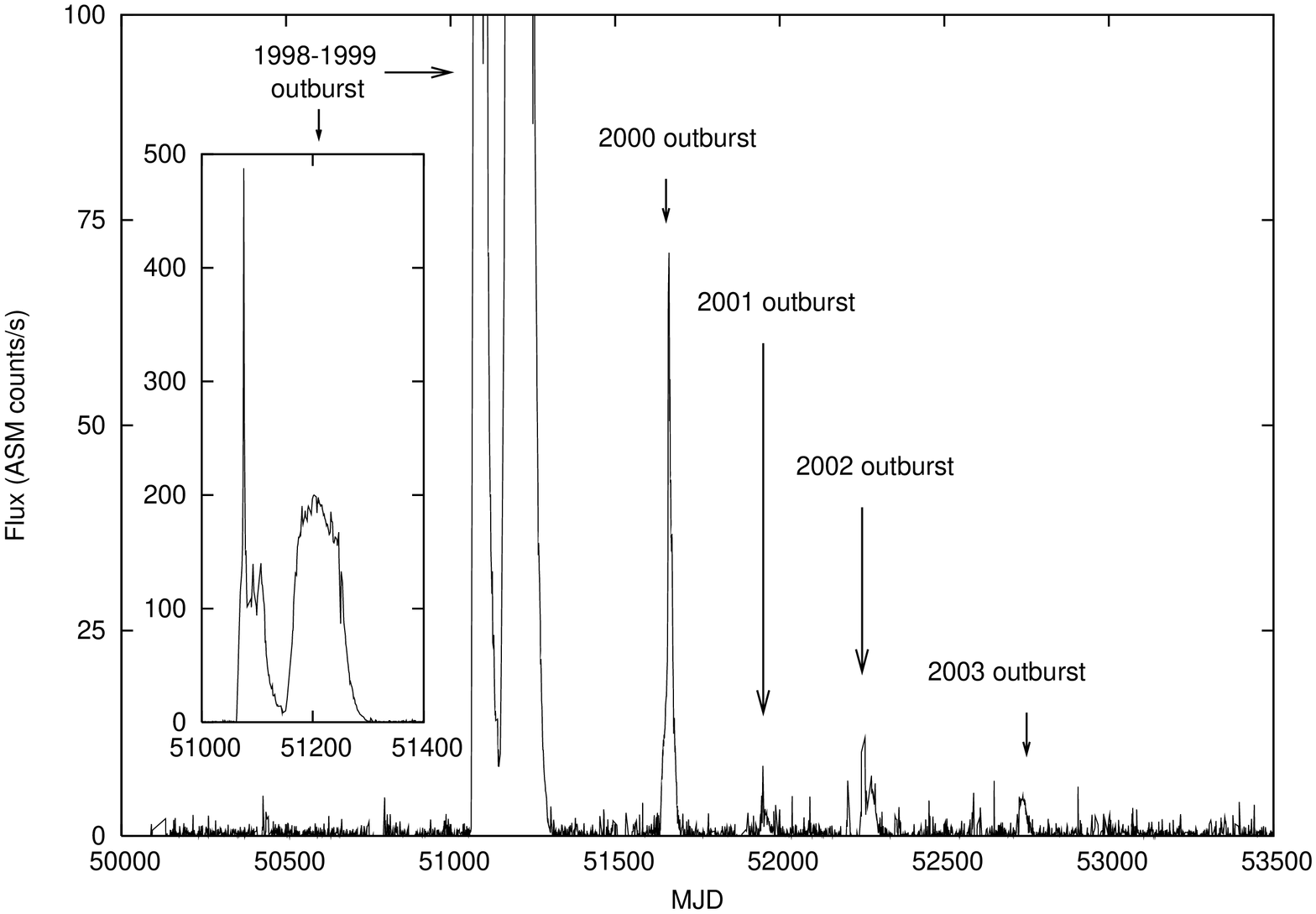}
        \includegraphics[angle=-90,width=9cm]{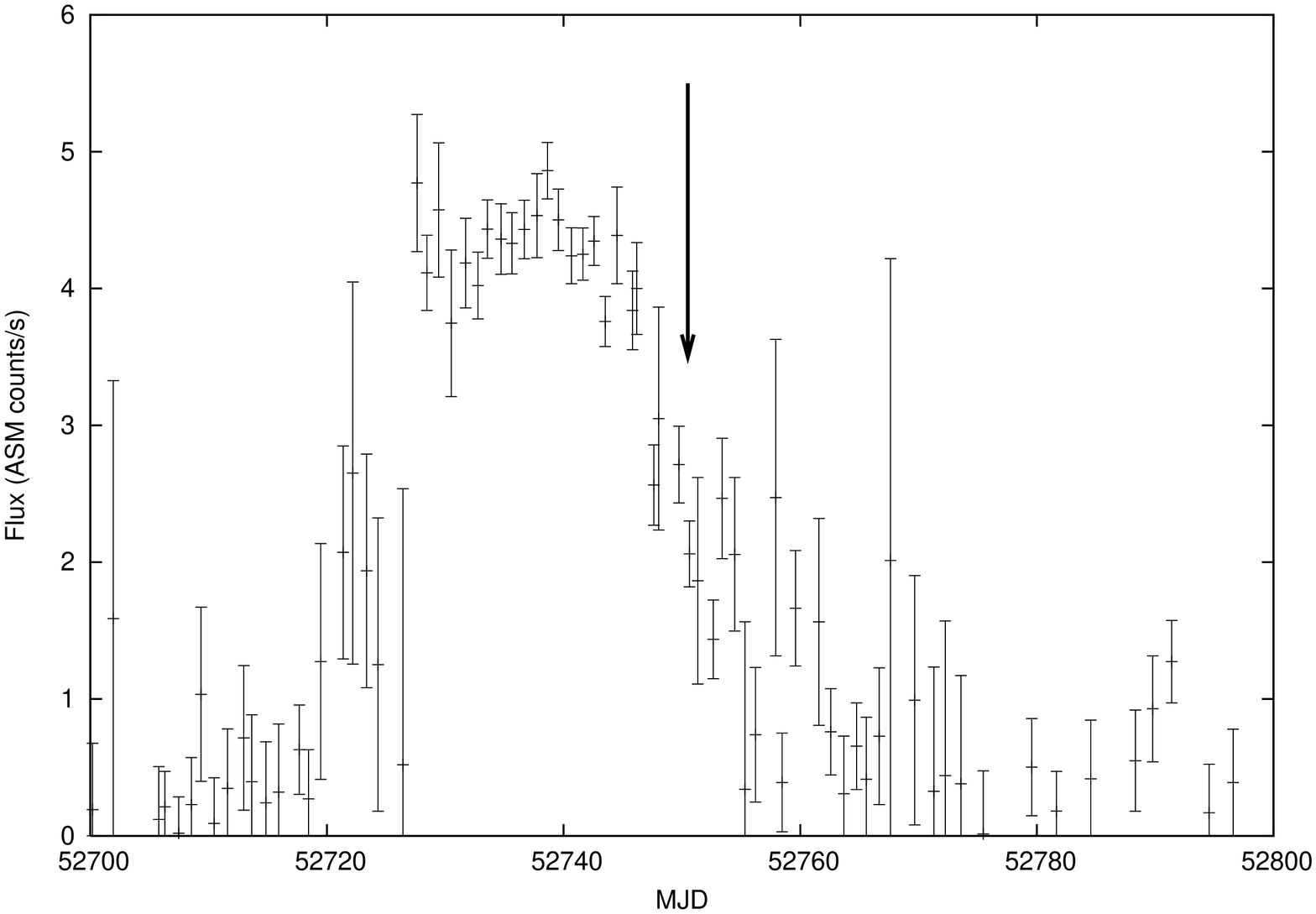}
	\end{center}
	\caption{Lightcurves of the flux of the microquasar $\xtejqcq$ in the 2-10\,keV band acquired by ASM/Rossi-XTE. Left: Lightcurve from 1996 to 2005.
          Right: zoom during the 2003 outburst. The date of our observations (2003 April 21) is marked by the arrow.}
	\label{ASM_overview}
	\end{figure*}

$\xtejqcq$ was discovered as a transient X-ray binary in September
1998 by the All Sky Monitor (ASM) onboard the {\it Rossi-XTE} satellite
\citep{remillard:1998}. Optical
\citep{orosz:1998,jain:1999} and radio \citep{campbell-wilson:1998}
counterparts were promptly identified, classifying $\xtejqcq$ as a
microquasar. The companion star is in a 1.541 day orbit \citep{jain:2001a}. It has been 
shown spectroscopically to be a G8\,IV--K4\,III  at a distance of 5.3\,kpc; the compact object, with a mass of $10.5 \pm 1.0 \Msol$,
is a black hole candidate \citep{orosz:2002}.  Five main
outbursts have been observed in $\xtejqcq$ (see Figure
\ref{ASM_overview} left part). The first one, which started in September 1998
and lasted about 200 days, was the most powerful: at its maximum, the
2-10 keV flux reached almost $500$\,ASM counts/s (=6.8\,Crabs). During
this outburst, the source had a complex behaviour and transited
through all canonical X-ray spectral states \citep{homan:2001}. 
In 2000, the source exhibited another
outburst, the 2-10 keV flux reaching around 1 Crab, and the source
transiting again through different spectral states. In 2001, 2002, and
2003, the source showed three less powerful outbursts, during which it
remained in the low/hard state \citep{arefev:2004,sturner:2005}. Such ``mini'' outbursts following a major eruption have been seen in other X-ray binaries (e.g. \citealt{simon:2010}) and in some dwarf novae \citep{kuulkers:1996}. The mechanism triggering these mini-outbursts is not understood within the framework of the standard LMXB disk instability model \citep{dubus:2001b}. They could be related to the accretion disk becoming eccentric during outburst in binary systems with small mass ratios \citep{hellier:2001}.

Overlapping contributions from the companion star, the accretion disk, and from the relativistic jet make the optical to near-infrared (NIR)  wavelength range particularly important to study. For instance, a compilation of optical and NIR observations of the black hole candidate X-ray binary GX\,339-4 showed that its NIR emission during low/hard states was non-thermal, likely synchrotron radiation emanating from the compact jet of this microquasar \citep{corbel:2002a}. Its spectral energy distribution (SED) showed the typical signature of a compact jet, namely a clear change in slope in the optical-NIR domain with an inverted power-law at lower frequencies. Similar signatures have been seen in other LMXBs \citep{kalemci:2005,migliari:2006}.

Here, we report on the results of optical and NIR observations of the microquasar $\xtejqcq$ during the 2003 outburst. 
$\xtejqcq$ remained in the low-hard state during the outburst.
Our goal was to assess whether emission from the compact jet extended to the NIR. The weakness of the X-ray ouburst made it potentially more favourable to detect a jet contribution, because the accretion luminosity is expected to decrease faster than the jet luminosity  \citep{heinz:2003}.
The optical and NIR observations and data reduction are described in Sect.~\ref{observations}. The results are presented in Sect.~\ref{results} and discussed in Sect.~\ref{conclusions}.

%%%%%%%%%%%%%%%%%%%%%%%%%%%%%%%%%%%%%%%%%%%%%%%%%%%%%%%%%%%%%%%
%%%%%%%%%%%%%%%%%%%%%%%%%%%%%%%%%%%%%%%%%%%%%%%%%%%%%%%%%%%%%%%

%__________________________________________________________________

\section{Observations and data reduction} \label{observations}

\subsection{Photometry and spectroscopy}

Our observations were performed during the night between 2003 April $21^{st}$ and $22^{nd}$ (Figure \ref{ASM_overview} right
shows when our observations took place during the X-ray
outburst), using the NTT (New Technology Telescope) telescope on the
La Silla observatory of ESO (European Southern Observatory), as part
of a Target of Opportunity (ToO) programme (PI S. Chaty). They took place on the
declining phase of the 2003 outburst of $\xtejqcq$ and consist of optical and NIR photometry (both deep and rapid), NIR spectroscopy and polarimetry. The results from the NIR polarimetry were reported in \cite{dubus:2006b}.

The NIR data were obtained in the J, H, and K$_s$ filters with the
spectro-imager SoFI (Son oF Isaac), using the large field imaging
(field of view of $4\aminp92 \times 4\aminp92$ and image scale of $0
\asecp 288$/pixel).  For each filter of the NIR deep photometric
observations, we observed the source at nine different positions with
60~s exposure time each, to estimate and substract the thermal sky
emission, with a standard shift-and-combine jitter procedure.  We also
performed rapid photometry in the K$_{\rm s}$ filter, observing the source
at different positions for three hours, with an integration time of 2~s
for each exposure. The readout mode was double correlated read, leading to an overhead time of $\sim$\,50\%; therefore we reached a time resolution of nearly 3~s for this rapid photometry.

To calibrate the photometric observations, we
observed two photometric standard stars of the Persson catalogue
\citep{persson:1998}: sj\,9136 and sj\,9146. Concerning the NIR
spectroscopy, we took 24 spectra of 60~s each, half with the blue grism
($0.95-1.64 \microns$) and half with the red grism ($1.53-2.52
\microns$); we also took spectra of the telluric standard Hip\,63689 to
correct the spectra of $\xtejqcq$ from the atmospheric absorption.

The optical data were obtained in the B, V, R, I and Z filters with
the spectro-imager EMMI (Extraordinaire Multi-Mode Imager), using the
large field imaging (field of view of $9\aminp9 \times 9\aminp1$,
binning $2 \times 2$ for a better sensitivity and image scale of $0
\asecp 332$/pixel).  We observed the source for 300~s in the B-band,
for 3~s and 60~s in the V-band and for 60~s in the R-, I- and Z-band. We
also observed the photometric standard stars PG\,1633 and PG\,1657. We
performed rapid photometry in the V filter (a series of 10~s exposures
during two hours). Concerning the optical spectroscopy, we took three
spectra of 300~s each and a spectrum of a spectro-photometric standard
star (LTT\,7379).

We used the Image Reduction and Analysis Facility (IRAF) suite to
perform data reduction, carrying out standard procedures of optical
and NIR image reduction, including flat-fielding and NIR sky
subtraction. For the standard photometry, we used a median filter
before carrying out aperture photometry with the {\it noao.daophot}
package. We obtained the apparent magnitudes $m_{app}$ from the
instrumental magnitudes $m_{inst}$ through the following formula,
where $Z_p$ is the zero-point, $ext$ the extinction coefficient, and
$airmass$ the airmass at the time of the observations:
$m_{app} = m_{inst} - Z_p - ext \times airmass$

We used the characteristic extinction coefficients at La Silla and
obtained the zero points given in Table \ref{coeff-ext-Zp} by
averaging the values obtained with the different standard stars. For
the Z filter, the apparent magnitudes of the standard stars were not
available, so we did not calibrate m$_{app}$(Z).

Concerning the spectra, we used the IRAF {\it noao.twodspec} package
to extract the spectra and perform wavelength calibration. Since we
did not observe any spectro-photometric standard stars, we did not
perform flux calibration. We divided the NIR spectra of $\xtejqcq$ by
the spectra of the telluric standard and multiplied it by the spectra
of a 4600~K blackbody (we took the average of the effective
temperature range of $\xtejqcq$ companion star given by
\citeauthor{orosz:2002} \citeyear{orosz:2002}: between 4100\,K and
5100\,K). We point out that the signal-to-noise ratio of the
optical spectra of $\xtejqcq$ was too faint to securely identify any
feature. Fortunately the NIR spectra were more exploitable, but not in the whole waveband coverage though, due to absorption.

\begin{figure}[!ht]
	\begin{center}
	\includegraphics[angle=-90,width=8.5cm]{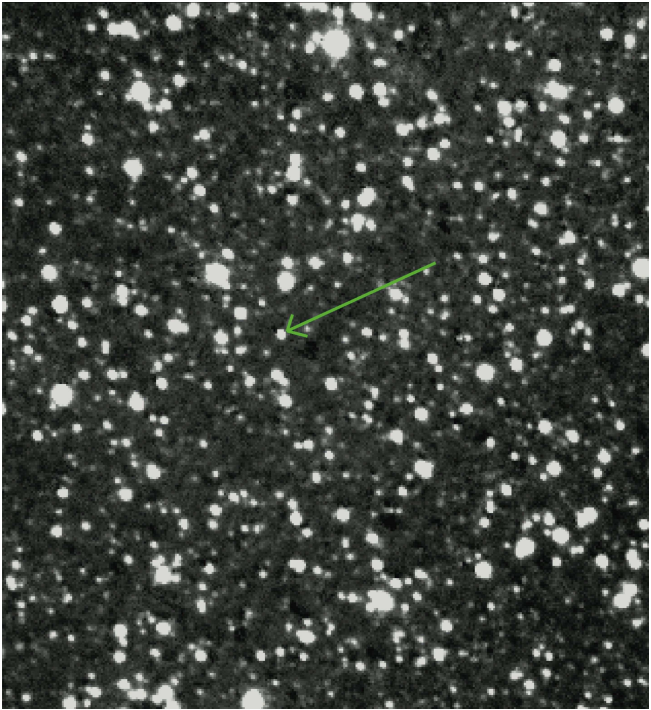}
	\end{center}
	\caption{$\xtejqcq$ field of view ($3.9' \times 3.9'$, H filter):
North is to the top and East to the left. The $\xtejqcq$ counterpart
is indicated by the arrow.}
	\label{champ_xte}
	\end{figure}

We give in Figure \ref{champ_xte} a finding chart in NIR wavelengths for $\xtejqcq$.

\subsection{Polarimetry} \label{section:polarimetry}

We also performed polarimetric observations on August 1 and 2, 2007. We collected a series of 10 second K$_s$ band exposures of the field around $\xtejqcq$ at the ESO NTT using SOFI in polarimetric mode. A Wollaston prism splits the incoming light into two images with perpendicular polarisation. $\xtejqcq$ was jittered along the mask for sky subtraction. Images taken at four different angles were used to compensate for the instrumental polarisation and this was checked against observations of unpolarised standards (see \citealt{dubus:2008} for details).

In addition, we also used polarimetric observations to derive the K$_{\rm s}$ magnitude in quiescence. Since we did not observe any photometric standard star during this run, we performed relative photometric calibration, thanks to isolated and bright stars of the 2MASS catalogue \citep{cutri:2003}, close to $\xtejqcq$. We obtained the following apparent magnitude for $\xtejqcq$: K$_s = 16.25 \pm 0.05 \mags$.

%%%%%%%%%%%%%%%%%%%%%%%%%%%%%%%%%%%%%%%%%%%%%%%%%%%%%%%%%%%%%%%
\begin{table}[!ht]
\caption{Characteristic extinction coefficients at La Silla and
derived zero-points for the different optical and NIR bands.}
	\begin{center}
	\begin{tabular}{c c c}
	\hline
	\hline
	Filter & ext. & Z$_{p}$ \\
	\hline
 	 B & 0.214 & 0.069 $\pm$ 0.002 \\
 	 V & 0.125 & -0.584 $\pm$ 0.002 \\
 	 R & 0.091 & -0.741 $\pm$ 0.003 \\
 	 I & 0.051 & -0.231 $\pm$ 0.003 \\
 	 J & 0.08  & 2.062 $\pm$ 0.010 \\
 	 H & 0.03  & 2.232 $\pm$ 0.006 \\
 	 K$_{s}$ & 0.05 & 2.799 $\pm$ 0.008\\
	\hline
	\end{tabular}
	\end{center}
	\label{coeff-ext-Zp}
\end{table}
%%%%%%%%%%%%%%%%%%%%%%%%%%%%%%%%%%%%%%%%%%%%%%%%%%%%%%%%%%%%%%%

%%%%%%%%%%%%%%%%%%%%%%%%%%%%%%%%%%%%%%%%%%%%%%%%%%%%%%%%%%%%%%%
%%%%%%%%%%%%%%%%%%%%%%%%%%%%%%%%%%%%%%%%%%%%%%%%%%%%%%%%%%%%%%%

\section{Results} \label{results}

\subsection{Photometry and extinction} \label{photometry}

We have estimated the interstellar absorption using the column density on the line of sight derived from {\em Chandra} observations: $\nh = 0.88 \pm 0.1 \times 10^{22}\cmmoinsdeux$ \citep{corbel:2006}. This value is somewhat lower than, but still consistent with, the HI column density integrated through the whole Galaxy, given by both the Leiden/Argentine/Bonn ($\nh = 1.01 \times 10^{22} \cmmoinsdeux$) and Dickey \& Lockman ($\nh = 0.897 \times 10^{22}\cmmoinsdeux$) surveys. %We thereafter deduced 
The interstellar absorption in the V-band A$_V$ is then deduced from the relation $\Av = 5.59 \times 10^{-22} \nh$ \citep{predehl:1995} and the different A$_\lambda$ using the relations established by \citet{cardelli:1989}. Table \ref{magapp} lists the apparent magnitudes we derived from our observations, together with the values obtained for A$_\lambda$ (taking into account the 1.6 $\sigma$ uncertainty on $\nh$) and, finally, the dereddened apparent magnitudes (taking into account the uncertainty on A$_\lambda$). The observed fluxes before and after corrections are plotted in the left part of Fig.~\ref{sed_2003}.

There is clearly a change of slope in between the NIR and visible wavelengths. The reddened NIR and optical spectral slopes are consistent with powerlaws of spectral index 1.7 and 3.9 respectively, and the dereddened NIR and optical with spectral index of 0.3 and -1.3 respectively\footnote{The spectral index $\alpha$ is defined as $F_{\nu} \propto \nu^{-\alpha}$}. The best-fitting powerlaws are indicated in Figure \ref{sed_2003} (left) by dotted lines.

In order for both the NIR and optical slopes to be roughly compatible, one would have to decrease the value of the column density to A$_V$=3.5 (corresponding to $\nh \sim 0.6 \times 10^{22} \cmmoinsdeux$), which is clearly well below the value derived from X-ray observations, and also from the HI surveys, including uncertainty. And even by doing this, the optical I magnitude point is never aligned with the NIR and optical slopes, confirming that a change of slope is clearly present between the optical and NIR power laws, with two different spectral indices. Because the NIR spectrum is optically thin with a positive spectral index, the jet break must be located either in the NIR (around the H-band as suggested by right panel of Fig. \ref{sed_2003}), or towards longer wavelengths, in the MIR domain. Only contemporaneous observations from optical to MIR domain would allow us to constrain the exact location of the jet break.

%%%%%%%%%%%%%%%%%%%%%%%%%%%%%%%%%%%%%%%%%%%%%%%%%%%%%%%%%%%%%%%
\begin{table}[!ht]
\caption{Apparent magnitudes, interstellar absorption, and
dereddened apparent magnitudes for various wavelengths.}
	\begin{center}
	\begin{tabular}{c c c c}

	\hline
	\hline
	Filter & m$_{app}$ & A$_\lambda$ & m$_{app}$-A$_\lambda$ \\
	
	\hline
 	 B & 20.00 $\pm$ 0.04 & 6.57 $\pm$ 0.75 & 13.43 $\pm$ 0.79 \\
 	 V & 18.48 $\pm$ 0.03 & 4.92 $\pm$ 0.56 & 13.56 $\pm$ 0.59 \\
 	 R & 17.35 $\pm$ 0.01 & 3.69 $\pm$ 0.42 & 13.56 $\pm$ 0.43 \\
 	 I & 16.33 $\pm$ 0.01 & 2.36 $\pm$ 0.27 & 13.97 $\pm$ 0.28 \\
 	 J & 14.50 $\pm$ 0.01 & 1.39 $\pm$ 0.16 & 13.11 $\pm$ 0.17 \\
 	 H & 13.46 $\pm$ 0.02 & 0.94 $\pm$ 0.11 & 12.52 $\pm$ 0.13 \\
    K$_{s}$ & 12.40 $\pm$ 0.01 & 0.56 $\pm$ 0.06 & 11.84 $\pm$ 0.07\\
	\hline

	\end{tabular}
	\end{center}
	\label{magapp}
\end{table}
%%%%%%%%%%%%%%%%%%%%%%%%%%%%%%%%%%%%%%%%%%%%%%%%%%%%%%%%%%%%%%%

\subsection{SED} \label{SED}

\begin{figure*}[!ht]
  \begin{center}
    \includegraphics[width=6.5cm,angle=90]{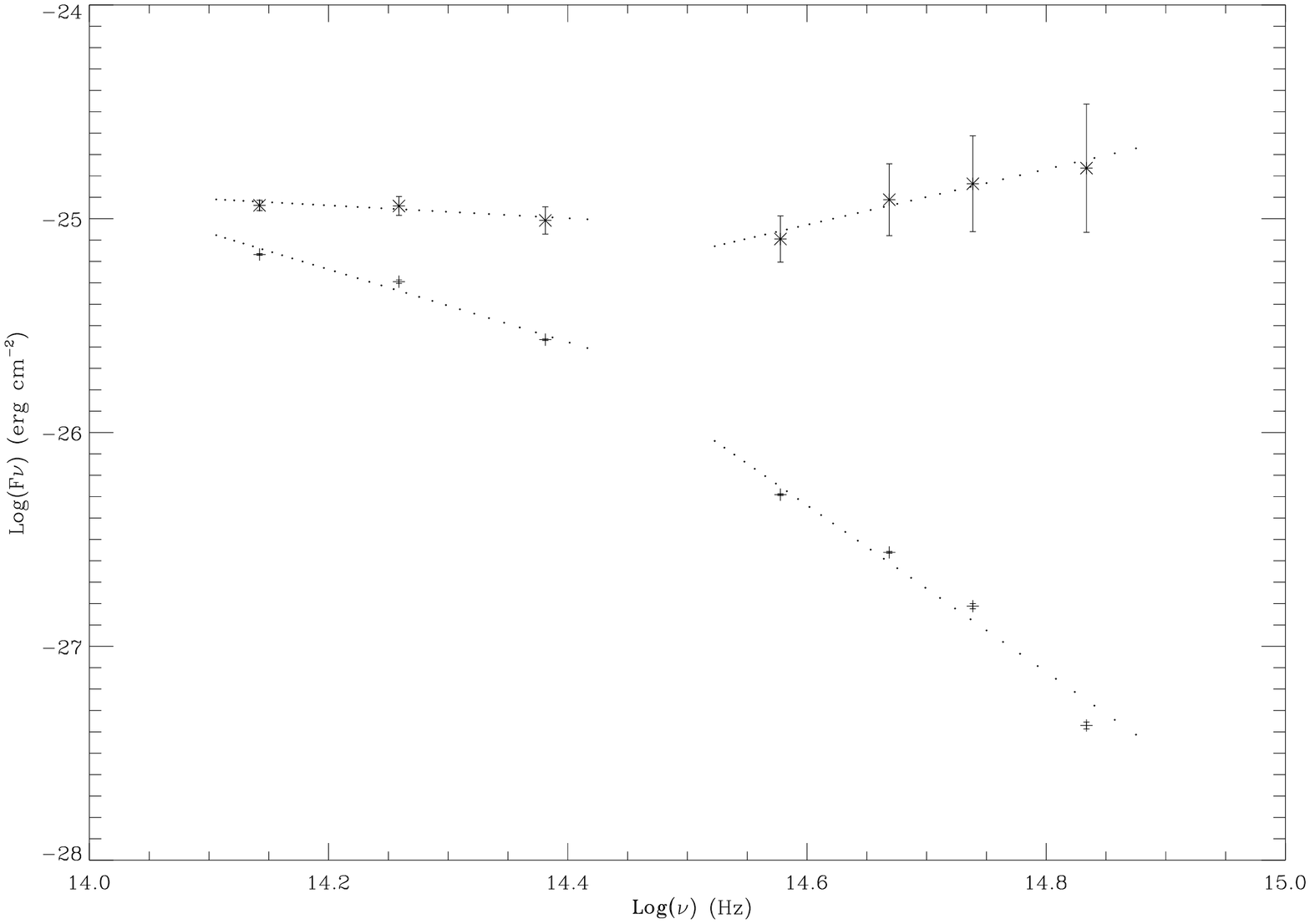}
    \includegraphics[width=6.5cm,angle=90]{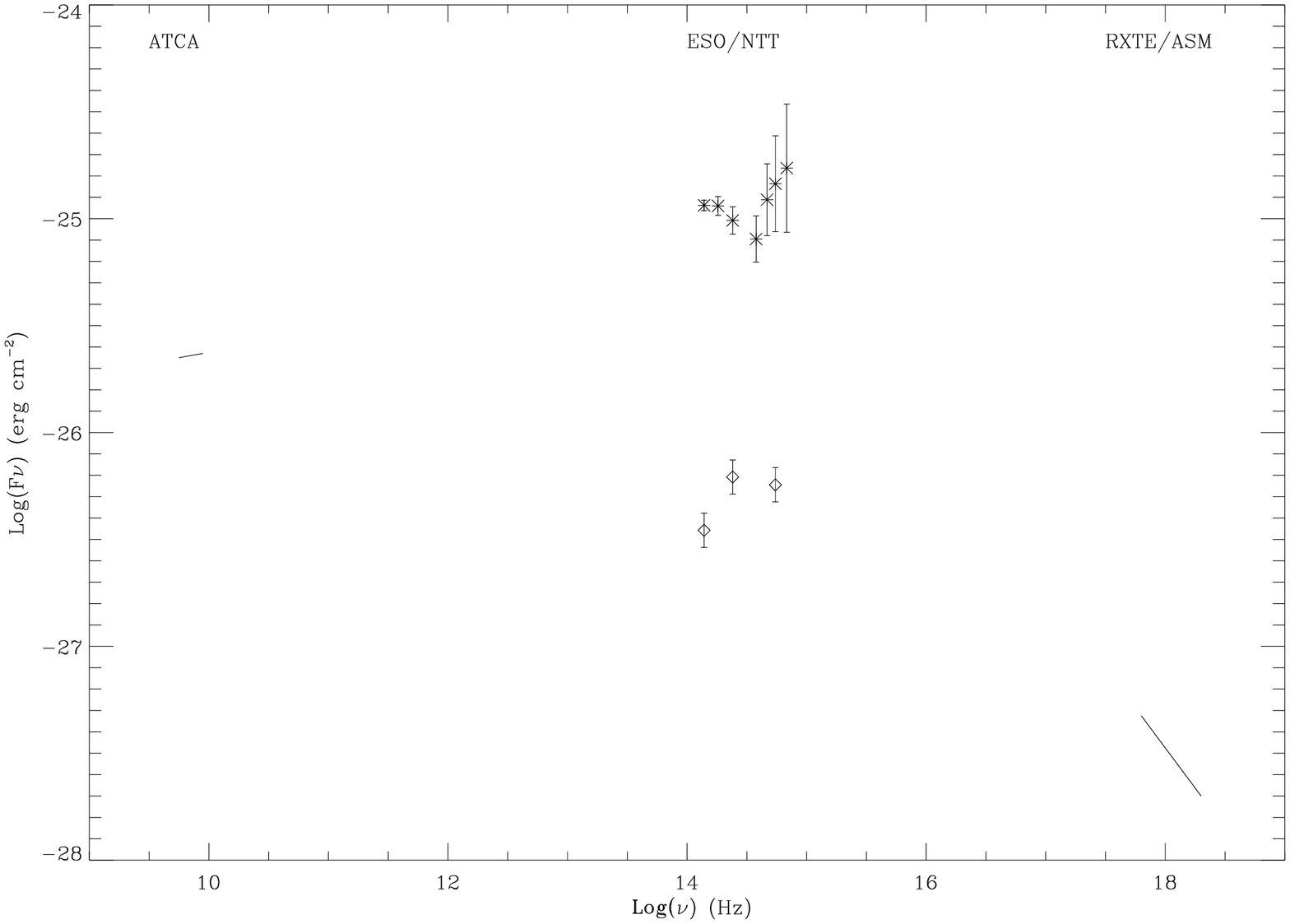}
  \end{center} %height=\linewidth
  \caption{$\xtejqcq$ SEDs during the 2003 mini-outburst. 
    Left: comparison between reddened (crosses in the lower part) and dereddened (asterisks in the upper part) optical/NIR SEDs. From optical (right) to IR (left): BVRI JHK$_{\rm s}$ points. The error bars are also shown and are bigger in the case of the dereddened points owing to the $1.6 \sigma$ uncertainty on the $\nh$ and then on A$_v$ (see section \ref{photometry}). We also show here power law slopes, with spectral indices given in the text.
Right: Broadband SED. Our ESO/NTT observations taken during the mini-outburst are shown with asterisk symbols, 
and the data taken during quiescence (K$_{\rm s}$ flux obtained during our polarimetric run, J and V flux from \citeauthor{orosz:2011} \citeyear{orosz:2011}) are shown with lozenges.
    All these data are dereddened with A$_V$ = 4.92 magnitudes, and the error bars show the uncertainty on the column density.
    We also plot the simultaneous X-ray flux/spectrum \citep{arefev:2004},
    and the ATCA radio fluxes observed during the 2002 outburst, with similar X-ray flux.
}
  \label{sed_2003}
\end{figure*}

The broad-band SED of $\xtejqcq$ from radio to X-rays is shown in the right part of Fig. \ref{sed_2003}, including our optical/NIR observations taken during the mini-outburst and also during quiescence. The optical/NIR data are dereddened from interstellar absorption.  The line in the lower right indicates the flux and spectral index of the simultaneous X-ray data of $\xtejqcq$ during the 2003 mini-outburst, as observed by ASM/{\it Rossi-XTE} \citep{arefev:2004}. We also include radio data obtained during the 2002 mini-outburst with a similar X-ray flux, reported with the line in the left. Although no contemporary radio observations could be found, this is indicative of the radio emission that could have been expected from the compact jet during the low/hard X-ray mini-outburst of 2003 (see Sect.~\ref{introduction}).

The optical data are consistent with the Rayleigh-Jeans tail of a multicolour
blackbody, which is characteristic of the emission coming from the outer part
of the accretion disk, whereas the NIR data suggest a non
thermal, inverted spectra, which is characteristic of synchrotron emission and is
usually associated with a compact radio jet 
(see e.g. \citeauthor{corbel:2002a} \citeyear{corbel:2002a} for $\gx$
and \citeauthor{chaty:2003b} \citeyear{chaty:2003b} for $\xtejodh$).  
The level of NIR emission is consistent with the extrapolation to high frequencies of the (non-contemporary) flat/inverted radio spectrum and the extrapolation to low frequencies of the X-ray spectrum (with photon index $\approx 1.6$, \citealt{arefev:2004, sturner:2005}).

%%%%%%%%%%%%%%%%%%%%%%%%%%%%%%%%%%%%%%%%%%%%%%%%%%%%%%%%%%%%%%%	
%%%%%%%%%%%%%%%%%%%%%%%%%%%%%%%%%%%%%%%%%%%%%%%%%%%%%%%%%%%%%%% 

%%%%%%%%%%%%%%%%%%%%%%%%%%%%%%%%%%%%%%%%%%%%%%%%%%%%%%%%%%%%%%% 
%\begin{figure}[!ht]
%  \begin{center}
%  \end{center}
%  \caption{}
%  \label{sed_2003}
%\end{figure}
%%%%%%%%%%%%%%%%%%%%%%%%%%%%%%%%%%%%%%%%%%%%%%%%%%%%%%%%%%%%%%%

\subsection{Colour-magnitude diagrams} \label{CMD}

\begin{figure*}[!ht]
	\begin{center}
	\includegraphics[width=6.5cm,angle=90]{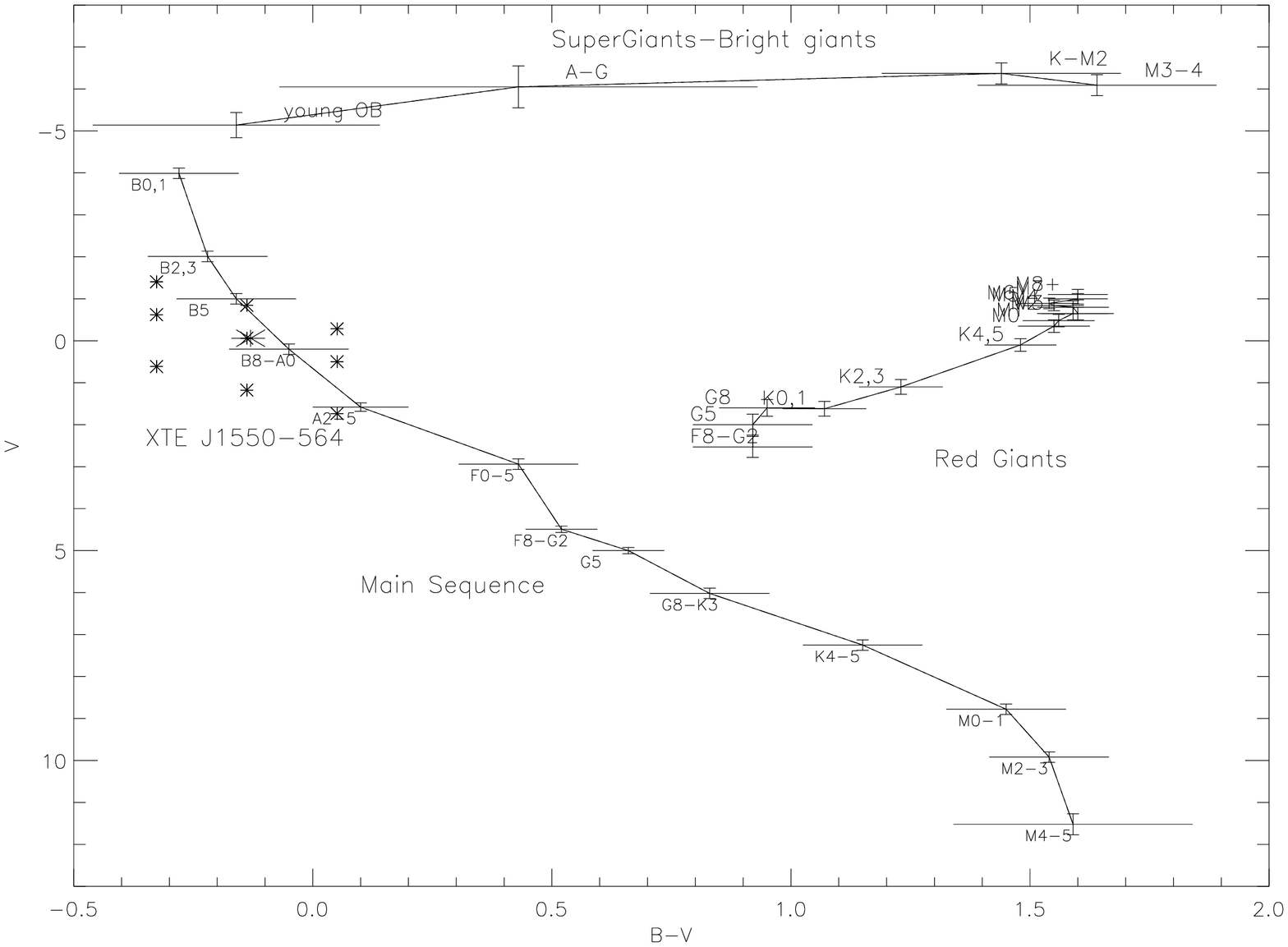}
	\includegraphics[width=6.5cm,angle=90]{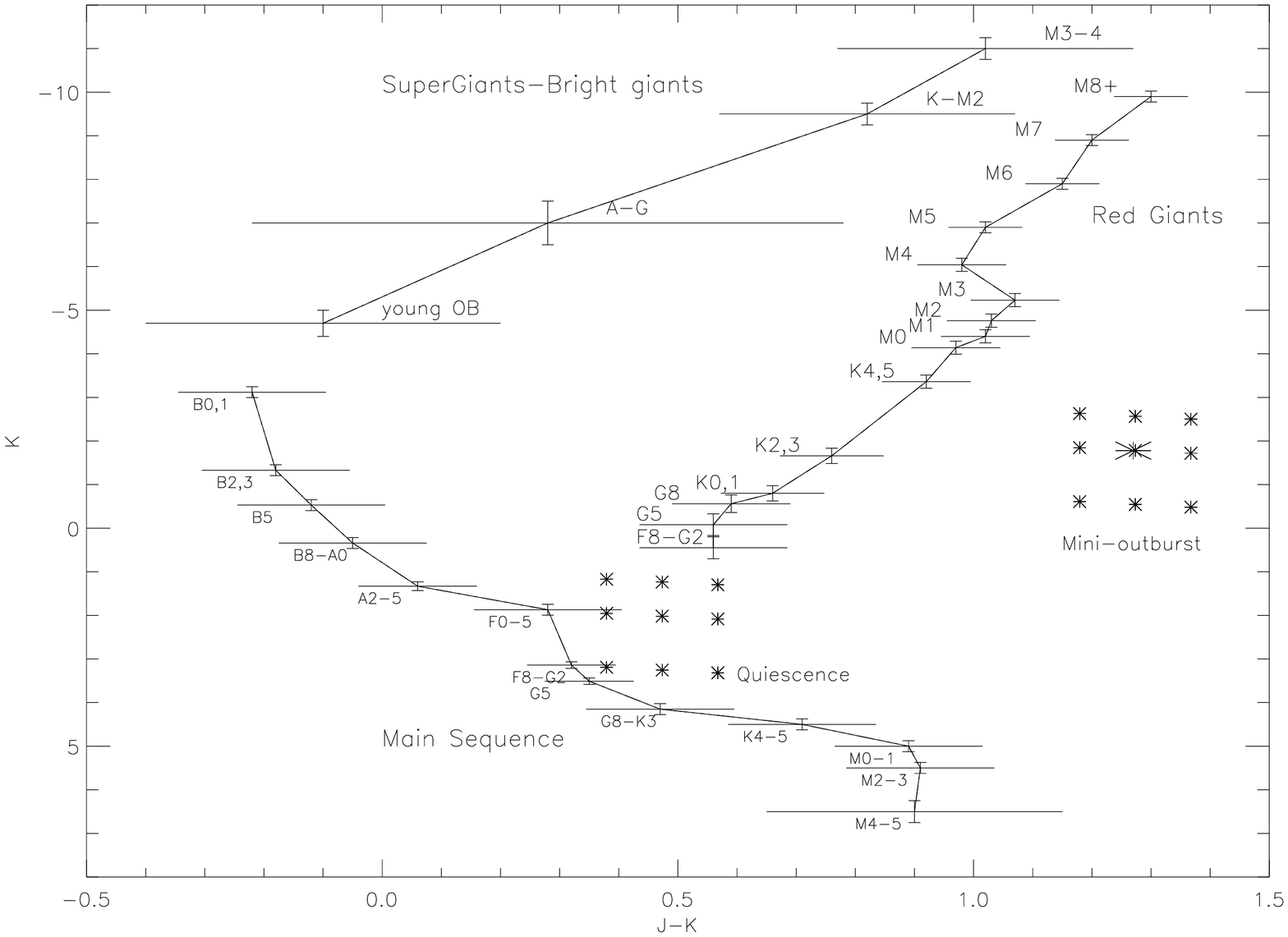}
	\end{center}
	\caption{CMDs with characteristic absolute magnitudes of
          various spectral types, on which we overplot our optical and NIR data (left and right plot respectively) for $\xtejqcq$ during the 2003 mini-outburst; 
          the nine asterisks represent the position of $\xtejqcq$ counterpart, taking into account the uncertainties on its distance and column density. The bigger central asterisk indicates its position using the best fitted distance and column density. In the NIR CMD we also add $\xtejqcq$ position corresponding to quiescence, using magnitudes from this paper and \cite{orosz:2011}.}
	\label{jk_k}
	\end{figure*}

The absolute magnitudes of $\xtejqcq$ during the mini-outburst are reported in the colour-magnitude diagrams (CMD) in Figure \ref{jk_k}. 
The absolute magnitude was computed via 
\[M_\lambda = m_\lambda + 5 - 5 \times \log(d(pc)) - A_\lambda.\]

In both CMDs, the big asterisk indicates the position of $\xtejqcq$
optical/NIR counterpart, and the small asterisks surrounding it represent the parameter space of the
source, taking into account the uncertainty on its distance ($d = 5.3
\pm 2.3 \kpc$; \citeauthor{orosz:2002} \citeyear{orosz:2002}) and on
the column density ($\nh = 0.88 \pm 0.1 \times 10^{22} \cmmoinsdeux$;
\citeauthor{corbel:2006} \citeyear{corbel:2006}, as reported in Table \ref{magapp}). According to
\citet{orosz:2002}, the companion star has a type from G8\,IV to K4\,III,
hence the quiescent state lies in the lower left part of the red
giant branch. We indicate in Figure \ref{jk_k} (right) the position of the $\xtejqcq$ quiescent magnitudes.

The NIR CMD (Fig. \ref{jk_k} right part) shows that the source is redder in outburst than in quiescence.
The companion star contributes more flux to the $J$ band than to the K$_{\rm s}$ band. Hence, the significantly redder colour implies an additional contribution with a flat or inverted spectrum.

The optical CMD (Fig. \ref{jk_k} left part) shows that the source was bluer in these bands in outburst compared to quiescence,
which is consistent with the dominant contribution of the accretion disk in the
optical. Indeed, when a multicolour blackbody spectrum is evolving
with a higher flux and temperature, the flux increases more in the
B-band than in the V-band, which implies a decrease in the (B-V) colour.

Both CMDs confirm that the optical and NIR fluxes evolve differently between the mini-outburst and quiescence.

%%%%%%%%%%%%%%%%%%%%%%%%%%%%%%%%%%%%%%%%%%%%%%%%%%%%%%%%%%%%%%%
%%%%%%%%%%%%%%%%%%%%%%%%%%%%%%%%%%%%%%%%%%%%%%%%%%%%%%%%%%%%%%%

\subsection{Rapid photometry} \label{rapidphot}

\begin{figure*}[!ht]
  \begin{center}
    \includegraphics[width=6.0cm,angle=-90]{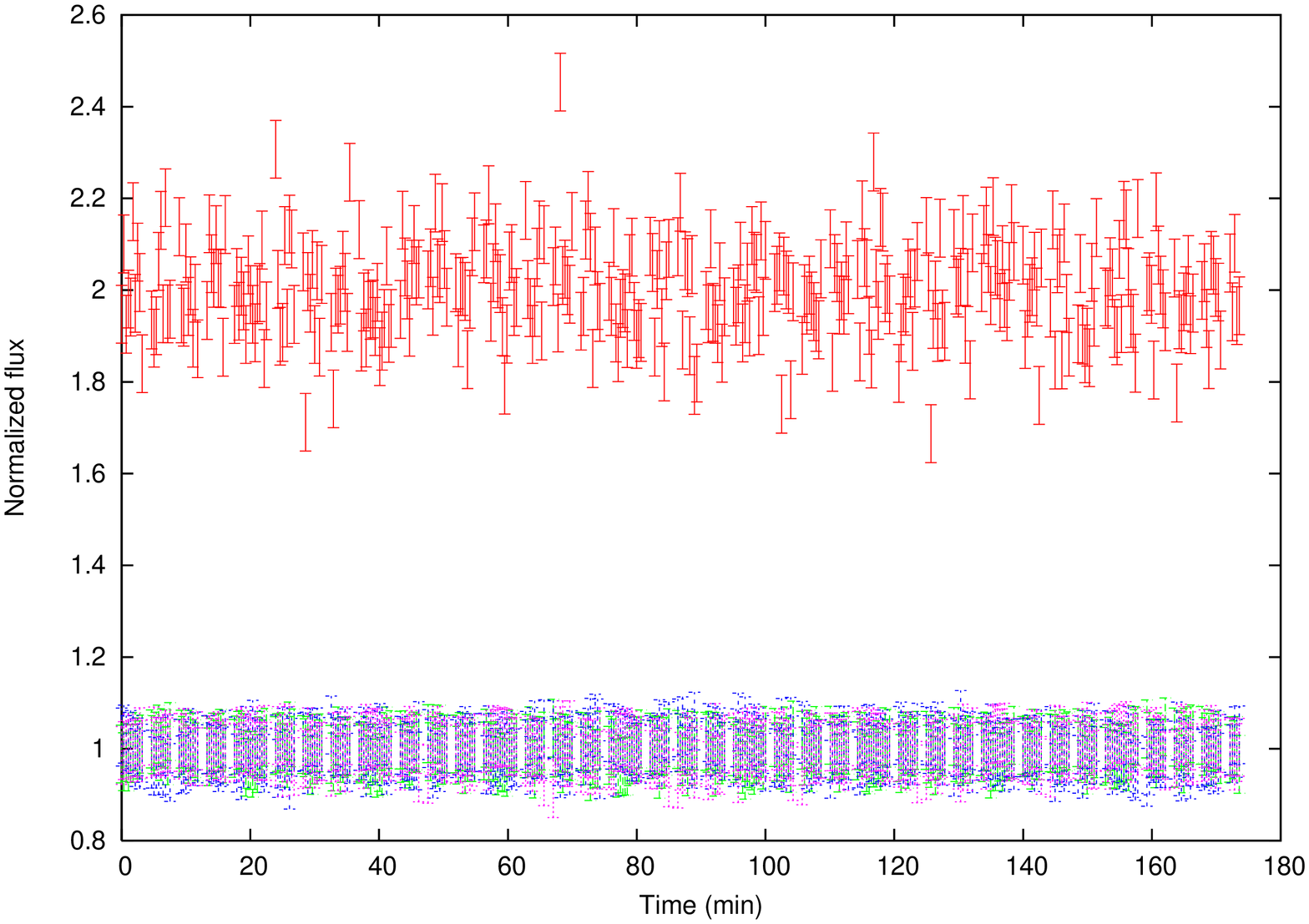}
    \includegraphics[width=6.0cm,angle=-90]{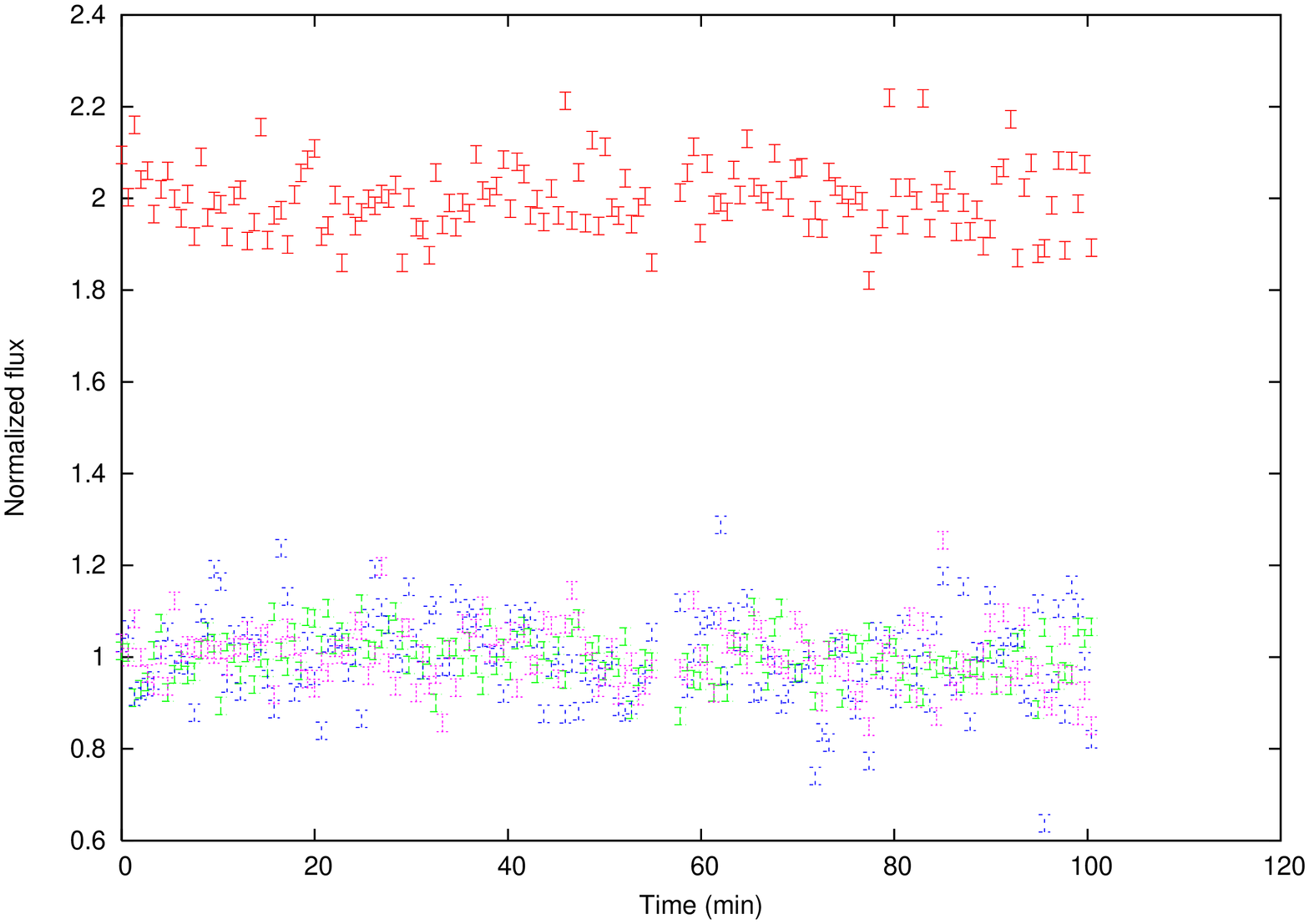}
  \end{center}
  \caption{Rapid photometry lighcurve in the K$_{s}$ filter (left) and  V filter (right): corrected and
    normalized flux of $\xtejqcq$ (red, artificially increased by 1 
    unit) and of stars of the field of view with similar brightness 
    (green, blue, purple respectively).}
  \label{phot_rpd_Ks}
\end{figure*}

We performed rapid photometry in the V and K$_{s}$ filters on $\xtejqcq$ and on
stars present in the field of view that were of comparable
brightness to $\xtejqcq$.  We then averaged the fluxes of these stars
to get a mean flux, and we finally divided the flux of $\xtejqcq$ by
this mean flux. These corrected and normalized fluxes are shown in
Figure \ref{phot_rpd_Ks} in the optical and NIR
respectively, where we can see the intrinsic variations of the X-ray
source and the surrounding stars. While rapid photometry in the NIR is rarely performed on LMXBs, it allows us to constrain rapid phenomena occuring on short timescale, either in the accretion disk, or related to the jet.

The rapid photometry shows that the source presents variations of
amplitude in the NIR greater than those of the surrounding stars, whereas in
the optical, the source behaviour is comparable with the
surrounding stars. The standard deviations of this optical and NIR
rapid photometry for $\xtejqcq$ and the surrounding stars are
presented in Table \ref{tab_phot_rpd}. 
Again, this suggests that the origin of the NIR and optical emission are different. However, note that variations in the optical of a similar amplitude to those in the NIR would probably not have been detected with these observations.

%%%%%%%%%%%%%%%%%%%%%%%%%%%%%%%%%%%%%%%%%%%%%%%%%%%%%%%%%%%%%%%
\begin{table}[!ht]
  \caption{Standard deviations taken from 
    the rapid photometry lightcurves of $\xtejqcq$ and of various stars from
    the field of view. 
    The stars have a brightness similar to the one of $\xtejqcq$.}
  \begin{center}
    \begin{tabular}{l c | l c}
      \hline
      \hline
      \multicolumn{2}{c}{K$_{s}$ filter} & \multicolumn{2}{c}{V filter}  \\
      \hline
      $\xtejqcq$ & 0.099 & $\xtejqcq$ & 0.074 \\
      Star \#1 & 0.020 & Star \#1 & 0.051 \\
      Star \#2 & 0.024 & Star \#2 & 0.065 \\
      Star \#3 & 0.021 & Star \#3 & 0.098 \\
      \hline
    \end{tabular}
  \end{center}
  \label{tab_phot_rpd}
\end{table}
%%%%%%%%%%%%%%%%%%%%%%%%%%%%%%%%%%%%%%%%%%%%%%%%%%%%%%%%%%%%%%%

%%%%%%%%%%%%%%%%%%%%%%%%%%%%%%%%%%%%%%%%%%%%%%%%%%%%%%%%%%%%%%%
%%%%%%%%%%%%%%%%%%%%%%%%%%%%%%%%%%%%%%%%%%%%%%%%%%%%%%%%%%%%%%%

\subsection{Spectroscopy} \label{spectroscopy}

\begin{figure}[!ht]
	\begin{center}
	\includegraphics[angle=0,width=9cm]{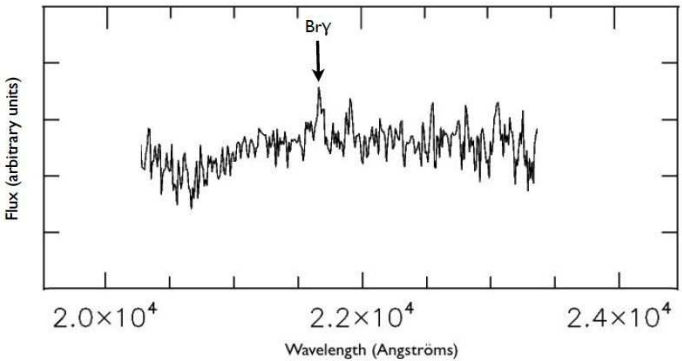}\hfill
	\end{center}
	\caption{Corrected NIR spectra of $\xtejqcq$ (the y-axis is in arbitrary units). Apart from the faint Br\,$\gamma$ emission line visible at 2.166\,$\microns$, both blue and red grisms are featureless.}
	\label{spectro_xte}
\end{figure}

We took two IR spectra in the blue and red grisms, both are very absorbed, with a low S/N ratio. As shown in Figure \ref{spectro_xte}, we only detect a faint emission line corresponding to Br\,$\gamma$ transition at 2.166\,$\microns$, very likely produced by the accretion disk. Apart from this, both NIR spectra are featureless, consistent with non-thermal emission emanating from the compact jet.
%{\bf *** Est-ce qu'on aurait pu detecter une raie br gamma ou autre venant d'un disque si celui-ci dominait ? Peut-on faire un commentaire la-dessus ?}

%%%%%%%%%%%%%%%%%%%%%%%%%%%%%%%%%%%%%%%%%%%%%%%%%%%%%%%%%%%%%%%%%%%%%%%%%%%
%%%%%%%%%%%%%%%%%%%%%%%%%%%%%%%%%%%%%%%%%%%%%%%%%%%%%%%%%%%%%%%%%%%%%%%%%%%

\subsection{Polarimetry} \label{polarimetry}

The NIR polarimetry of $\xtejqcq$, taken on the same night as part of the same observing programme, was reported earlier \citep{dubus:2006b}. The polarimetry showed an excess polarisation in $\xtejqcq$ compared with other stars in the field-of-view. Here, we report on polarimetric observations collected well after the mini-outburst described in this paper. The goal was to obtain a better measure of the polarisation signal from field stars in order to find the absolute value of the NIR polarisation from $\xtejqcq$ during its 2003 mini-outburst.

As described in Section \ref{section:polarimetry}, $\xtejqcq$ was detected in quiescence at a magnitude K$_s = 16.25 \pm 0.05 \mags$, but the low S/N prevented a meaningful polarisation measurement. The mean polarisation of the bright stars within $1\amin$ of $\xtejqcq$ (see right panel of Fig.~1 in \citeauthor{dubus:2006b} \citeyear{dubus:2006b}) was found to be $\approx 1.4\%$. The corresponding reduced Stokes parameter values are $q=-0.8\pm0.3$\% and $u=1.1\pm0.5$\% with the errors derived from the scatter of the stars $q$ and $u$ values. Assuming the polarisation of the field stars has not changed between 2003 and 2007, we correct for this mean polarisation and find that the absolute K$_s$ band polarisation during the 2003 mini-outburst of $\xtejqcq$ was about 2.4\%. This confirms that the polarisation fraction was in excess of the interstellar polarisation that could be expected with E(B-V)=0.7 ($\approx 0.7$\%, see \citealt{dubus:2006b}). We were unable to determine the angle correction. The few polarised standards available to this effect are very bright. Several were observed by defocusing the telescope, but the resulting photometry proved too unreliable to be of use. Hence, we cannot give the orientation of the polarisation angle with respect to the jet axis.

\section{Discussion} \label{conclusions}

\subsection{The spectral break}
We detected a break between the NIR and optical wavelengths during a mini-outburst of $\xtejqcq$, with a positive spectral index in NIR, suggesting optically thin emission consistent with a jet spectrum. The break is clearly visible in the spectral energy distribution and colour-magnitude diagrams. Inaccuracies in the column density used to deabsorb the fluxes cannot account for this break. The NIR spectrum is featureless, apart from a faint Br\,$\gamma$ emission line. The K$_{\rm s}$ lightcurve shows $\approx 10$\% variability on short timescales, which is suggestive of a non-thermal component. Evidence for an intrinsic IR polarisation during the outburst also points towards synchrotron emission \citep{dubus:2006b}. The overall SED is reminiscent of the low/hard state SED of GX 339-4 and XTE J1118-480 where the infrared emission was attributed to the compact jet (see \S1). Our data lead us to conclude that synchrotron jet emission dominated in the NIR during the 2003 mini-outburst of $\xtejqcq$.

A dominant jet contribution in NIR was also put forward by \cite{russell:2010} to explain the correlations between the $H$ band and X-ray fluxes during the 2000 outburst of $\xtejqcq$.
The 3-10 keV flux during our observations was about $6\times 10^{-10}$\ erg\ cm$^{-1}$ s$^{-1}$. Comparing our results with Fig.~1 of  \citet{russell:2010}, we find that with $H=13.46$ the source was in NIR about 50\% brighter during our observations than during the decline of the 2000 outburst, at the time when the source reached the same X-ray flux. Our observations are consistent with the picture of an increasing jet contribution in the NIR as the source becomes harder and fainter in X-rays \citep{russell:2006}.

The $V$ flux in the 2003 mini-outburst  is brighter than that seen in 2000 for similar X-ray luminosities \citep{jain:2001b,russell:2010}. The optical spectrum we measure is compatible with a Rayleigh-Jeans tail. This requires the temperature of the outer disk radius to be $\ga 10^4$~K. Otherwise, the flat part of the disk blackbody spectrum ($F_\nu\propto \nu^{1/3}$) should be visible\footnote{Steeper accretion disk spectra can be expected when the disk is irradiated \citep{hynes:2005} but this is unlikely to be the case here given the weak X-ray flux in outburst.}. Here, we assume the temperature distribution as a function of disk radius $R$ is $T_{\rm disk}\propto R^{-3/4}$, which is adequate for an accretion disk in outburst  \citep{dubus:2001b}. The temperature is high enough to ionize hydrogen in the outer disk, as expected in outburst. The data during the 2000 decline did not show a break in the spectra between $H$, $I$ and $V$ bands. The weaker $V$ flux in 2000 may have been due to a hotter disk temperature: this would have placed the Rayleigh-Jeans tail at higher frequencies.

\subsection{Physical conditions at the jet base}
The infrared fluxes decrease progressively from 12.0 mJy in $K$ to 6.6 mJy in $I$ before increasing again. The spectral steepening suggests the transition in the compact jet to optically thin synchrotron emission occurs at IR frequencies. The optically thin emission is dominated by the emission from the innermost region in self-absorbed jet models \citep{blandford:1979,hjellming:1988,kaiser:2006}. This is usually the case because the optically thin flux along the jet decreases rapidly with distance. The turnover frequency decreases along the jet and the summed contribution produces the flat spectrum at lower frequencies. 

Assuming the innermost region with cross-section radius $R_0$ and length $H_0$ is seen sideways, the transition from thick to thin synchrotron emission occurs at $\tau_\nu=\alpha_\nu R_0\approx 1$. The synchrotron emission and absorption coefficients have analytical expressions for a power-law distribution of electrons with an index $p$. Further assuming that the energy density in non thermal electrons is a fraction $\xi$ of the magnetic energy density $B_0^2/8\pi$ in the region, this gives a relationship between the peak frequency, $\xi$, $R_0$ and $B_0$. The approximate flux at the peak frequency can be derived and depends on $\xi$, $R_0$, $H_0$ and $B_0$. We find the following relationships for $B_0$ and $R_0$ (see Appendix A)
\begin{eqnarray}
B_0&\approx& 5\times 10^4\  \nu_{14} S_{10}^{-1/9} \xi^{-2/9} {h}^{1/9} d_{5}^{-2/9} {\rm \ G} \label{eq2}\\
R_0&\approx& 2.5\times 10^8\  \nu_{14}^{-1} S_{10}^{17/36} \xi^{-1/18} {h}^{-17/36} d_{5}^{17/18} {\rm \ cm}
\label{br}
\end{eqnarray}
where we have taken a minimum Lorentz factor for the electrons $\gamma_{\rm min}=1$, $p=2.5$ (optically thin spectral index of 0.75), $\nu_{\rm peak}=10^{14} \nu_{14}$\ Hz, $S_{\rm peak}=10\ S_{10}$\ mJy, $d=5\ d_5$\ kpc  and $H_0=h R_0$. Similar results are obtained for $p=2$ or $p=3$. These equations apply equally to black hole or neutron star LMXBs. The magnetic field depends most sensitively on the turnover frequency. 

Self-absorbed models usually make the assumption that adiabatic cooling is dominant over the radiative timescales. The adiabatic timescale is $t_{\rm ad}\ga R_0/c\approx 8$~ms. The synchrotron timescale is $t_{\rm sync}\propto \gamma^{-1} B^{-2}_0 \approx 120 \gamma^{-1}$~ms with the magnetic field derived above. Since NIR emission requires electron Lorentz factors $\gamma\approx 20$, it means that this is marginally verified at the jet base. We find variability on a few second timescales. $K$ band variability on $\approx 200$~ms timescales was also reported in GX 339-4 during a low/hard state \citep{casella:2010} but the above suggests that there could be NIR variability down to 10 ms timescales. Beyond the jet base,  $t_{\rm sync}/t_{\rm ad}$ increases rapidly because $B\sim z^{-1}$ along the jet axis $z$ whereas $R\sim z^\beta$ with $\beta\approx 0.5$ to reproduce flat spectra (e.g.  \citealt{hjellming:1988, kaiser:2006}). Self-Compton cooling can be ignored because a luminosity ratio of Compton to synchrotron emission $L_{\rm ic}/L_{\rm sync}\approx 0.2$ is inferred using the $R_0$ and $B_0$ given above. It can be shown that  $L_{\rm ic}/L_{\rm sync}\propto \nu_{\rm peak} S_{\rm peak}^{-5/18}$. 
Synchrotron self-Compton emission from the jet base will be negligible unless $\nu_{\rm peak}$ moves into the visible.

\citet{russell:2010} speculate that the X-ray emission in the hard state becomes fully jet-dominated when the 3-10 keV flux is below a few 10$^{-10}$ erg\ cm$^{-2}$\ s$^{-1}$. Our observations do show that the NIR emission lies close to the extrapolated X-ray spectrum.  If the cutoff at $\approx$ 100 keV in INTEGRAL is caused by synchrotron emission (but see \citealt{zdziarski:2004}), then the maximum electron Lorentz factor is $\gamma_{\rm max}\approx 6000$. The X-ray emitting electrons should be radiatively cooled and the X-ray spectrum below the cutoff should have a photon index $\ga 2$ when an index $\approx 1.7$ is observed.  

Finally, a magnetic field of a few teslas at the jet base is inevitable regardless of the detailed model if the NIR break is due to self-absorption. For comparison, the equipartition magnetic field with thermal pressure in the accretion disk at a radius close to the compact object is 
\begin{equation}
B_{\rm eq}\approx 5\times 10^7\  \eta^{-1/2} M_1^{-1/2} \dot{m}^{1/2} r^{-5/4} {\rm \ G,}
\end{equation}
where $r$ is the radius in units of the last stable orbit, $M_1$ is the mass of the compact object in solar masses, $\dot{m}$ is the accretion rate in units of Eddington and $\eta=H/R\rightarrow 1$ for Bondi-Hoyle or radiatively inefficient accretion. The X-ray luminosity during our observations is $\approx 10^{-3} L_{\rm Edd}$, which implies $\dot{m}\ga 10^{-3}$.  Therefore, the magnetic field at the jet base in $\xtejqcq$ represents at most 1\% of the equipartition magnetic field at the innermost radius.

Curiously, compact jets in AGNs typically have turnovers in the 1-100 GHz range \citep{kellermann:1981}, exactly as expected if the ratio $B/B_{\rm eq}$ were constant from microquasars to quasars. % ($B$ depends almost exclusively on $\nu$, Eq.~\ref{eq2}). 
If this ratio is also constant in an object, then the turnover frequency will move to longer wavelengths as the mass accretion rate decreases. This can be tested observationally in microquasars when they decline from outbursts.

\section{Conclusion}

We have obtained simultaneous NIR to optical coverage of the microquasar $\xtejqcq$ during a mini-outburst. Our dataset shows a  break in the SED from the NIR to the optical. The optical emission is compatible with the Rayleigh-Jeans tail of the accretion disk. The lack of prominent spectral feature in the NIR, the fast variability and the evidence for intrinsic polarisation lead us to attribute the NIR emission to synchrotron radiation from the compact jet. Based on correlations between IR and X-ray fluxes during its 2000 outburst, \citet{russell:2010} also interpreted the NIR emission from $\xtejqcq$ as jet emission. Evidence for NIR or optical jet emission from $\xtejqcq$ was also suggested by \citet{jain:2001b,corbel:2001,russell:2007b}. The NIR luminosity represents about 1.7\% of the X-ray luminosity. The jet contribution appears to be more important, in terms of the NIR to X-ray ratio, during the faint 2003 mini-outburst than during the 2000 outburst.

The SED shows a steepening from $K$ to $I$, suggesting the transition from optically thick to thin synchrotron emission occurs around 10$^{14}$ Hz. If this interpretation is correct, then the magnetic field at the jet base is at most a few teslas, or about 1\% of the equipartition magnetic field in the accretion disk close to the black hole as in AGN compact jets. The NIR emission region must be small and sub-second variability can be expected. 

Our data provide only a snapshot of the SED during an outburst. The evolution of the jet break during an outburst can provide important diagnostics of the jet physics \citep{heinz:2003,markoff:2003}. Good sampling of the optical to NIR SED both in time and frequency, ideally in combination with polarisation measurements, is required to identify this break independently of the radio or X-ray observations and to test models that suggest jet emission can dominate the X-ray emission.

\begin{acknowledgements}
  SC thanks the ESO staff for performing service observations, and SC and GD are grateful to an anonymous referee who helped to improve the paper.
  IRAF is distributed by the National Optical Astronomy Observatories, which are operated by the Association of Universities for Research in Astronomy, Inc., under a cooperative agreement with the National Science Foundation.
  {\it Rossi-XTE} Results were provided by the ASM/{\it Rossi-XTE} teams at MIT and at the {\it Rossi-XTE} SOF and GOF at NASA's GSFC. 
  This research has made use of NASA's Astrophysics Data System Bibliographic Services.  
  This publication makes use of data products from the Two Micron All Sky Survey, which is a joint project of the University of Massachusetts and the Infrared Processing and Analysis Center/California Institute of Technology, funded by the National Aeronautics and Space Administration and the National Science Foundation.
  This work was supported by the Centre National d'Etudes Spatiales (CNES), based on observations obtained with MINE --the Multi-wavelength INTEGRAL NEtwork--; and by the European Community via contract ERC-StG-200911.
\end{acknowledgements}

\appendix
\section{Magnetic field and size of the jet base}

We assume that the synchrotron-emitting region at the jet base is a  homogeneous cylinder of radius $R_0$ and height $H_0=h R_0$. The electrons follow a power-law distribution $dN=K_0\gamma^{-p}d\gamma$; the magnetic field is $B_0$. The standard formula for the synchrotron absorption coefficient  $\alpha_\nu$ is \citep{rybicki:1979}
\begin{equation}
\alpha_\nu=\frac{\sqrt{3}e^3}{8\pi m^2_e c^2}\left(\frac{3 e}{2\pi m_e c}\right)^{\frac{p}{2}} \Gamma\left(\frac{3p+2}{12}\right) \Gamma\left(\frac{3p+22}{12}\right)K_0 B_0^{\frac{p+2}{2}}  \nu^{-\frac{p+4}{2}}. 
\end{equation}
Similarly, the optically thin emissivity $j_\nu$ is 
\begin{equation}
j_\nu=\frac{\sqrt{3} e^3}{2\pi m_e c^2}\left(\frac{m_e c}{3e}\right)^{-\frac{p-1}{2}}\Gamma\left(\frac{3p+19}{12}\right)\Gamma\left(\frac{3p-1}{12}\right)  \frac{K_0 B_0^{\frac{p+1}{2}} }{p+1}\nu^{-\frac{p-1}{2}},
\end{equation}
so that the flux from the region can be written as $S_\nu=(1/2) (R_0/d)^2 H_0 j_\nu$, with $d$ the distance to the source. The synchrotron self-absorbed emission peaks at the frequency $\nu_{\rm peak}$ where  $\tau_\nu=\alpha_\nu R_0\approx 1$. The jet emission  transits from the flat optically thick part to the optically thin part at $\nu_{\rm peak}$, regardless of the detailed emission further down the jet (which only affects emission at frequencies below $\nu_{\rm peak}$). Inverting $S_\nu=S_{\rm peak}$ and $\tau_{\rm peak}=1$ gives two equations on $R_0$ and $B_0$, as functions of $S_{\rm peak}$, $\nu_{\rm peak}$, $h$, $p$ and $K_0$. We assume the energy in non-thermal electrons $\epsilon_e$  is a fraction $\xi$ of the magnetic field energy density
\begin{equation}
\epsilon_e\equiv \int_{\gamma_{\rm min}}^{\gamma_{\rm max}} K_0  \gamma^{1-p}m_e c^2 d\gamma=  \xi \frac{B_0^2}{8\pi}.
\end{equation}
$K_0$ can be expressed as a function of the equipartition fraction $\xi$, $B_0$, $p$, $\gamma_{\rm min}$ and $\gamma_{\rm max}$. We have assumed $p=2.5$ and $\gamma_{\rm max}\gg \gamma_{\rm min}=1$ in deriving Eq.~\ref{br}.

\end{document}